\documentclass[12pt]{article}
\usepackage{graphicx}

\begin{document}

\begin{titlepage}
\vskip 0.3cm

\centerline{\large \bf POMERON MODELS AND EXCHANGE DEGENERACY}
\centerline{\large \bf OF THE REGGE TRAJECTORIES}

\vskip 0.7cm

\centerline{J. Kontros$^{\ast}$, K. Kontros$^{\dagger}$, A.
Lengyel$^{\ddagger}$}

\vskip .3cm

\centerline{\sl Institute of Electron Physics, Universitetska 21,}
\centerline{\sl 88000 Uzhgorod, Ukraine}

\vskip 1.5cm

\begin{abstract}
Two models for the Pomeron, supplemented by exchange-degenerate
sub-leading Regge trajectories, are fitted to the forward
scattering data for a number of reactions. By considering new
Pomeron models, we extend the recent results of the COMPAS group,
being consistent with our predecessors.
\end{abstract}

\vskip .3cm

\vskip 12cm

\hrule

\vskip 1cm

\noindent
\vfill
$ \begin{array}{ll}
^{\ast}\mbox{{\it e-mail address:}} &
 \mbox{jeno@kontr.uzhgorod.ua}
\end{array}
$

$ \begin{array}{ll}
^{\dagger}\mbox{{\it e-mail address:}} &
   \mbox{kornel@kontr.uzhgorod.ua}
\end{array}
$

$ \begin{array}{ll}
^{\ddagger}\mbox{{\it e-mail address:}} &
 \mbox{sasha@len.uzhgorod.ua}
\end{array}
$

\vfill
\end{titlepage}\eject
\baselineskip=14pt

\section{Introduction}

Ever since the discovery of the rising $K^{+}p$ and $\pi ^{+}p$
total cross sections in Protvino \cite{Denisov}, experimental data
on hadronic total cross sections are among the most reliable tools
for checking models of the Pomeron. The so-called ''enhanced''
cuts were among the first theoretical ideas used to explain the
phenomenon \cite{Ter,Barger}. Since then, both the theory and
experiment made enormous progress. Apart from hadronic reactions,
measured now at much higher energies, photon-induced reactions
also confirm the universal rise of the total cross sections.
Empirical fits searching a universal Pomeron fitting all of the
existing data were made e.g. in refs. \cite{DL,DGLM,Ezhela}.

The simplest models as the contribution to the total cross sections can be
classified in three groups:

the dipole Pomeron: $\sigma \sim \log s$ \cite{JMS,Leader};

the ''Froissart model'': $\sigma \sim \log ^2s$ \cite{SS,BW};

the ''supercritical'' Pomeron: $\sigma \sim s^\varepsilon$
\cite{CW,DM,DL2} (actually, the list of relevant references is
much longer and we apologize to those omitted there).

The last option, the supercritical one, in QCD corresponds to an infinite
ladder of reggeized gluon exchange \cite{BFKL}. At finite energies, however
only a finite number of diagrams contributes, giving rise to a finite series
in logarithms of $s$ \cite{FJKLPP} like
\begin{equation}
\sigma \sim \sigma _0+\sigma _1\log s+\sigma _2\log ^2s,  \label{eq1}
\end{equation}
respecting the Froissart bound. In the language of Regge-poles such a finite
series corresponds to multipoles \cite{DJ}.

On the other hand, an obvious way to generalize \cite{DGLM} the
Donnachie-Landshoff model \cite{DL2} is to add another constant,
whenafter it becomes
\begin{equation}
\sigma \sim \sigma _0+\sigma _1s^\varepsilon .  \label{eq2}
\end{equation}
In ref. \cite{GN} the model (\ref{eq2}) was interpreted as that
corresponding to two Pomeron poles.

Recently a detail analysis of three Pomeron models, namely
\begin{equation}
\mbox{I. }P\sim \sigma _0s^\varepsilon ,  \label{eq3}
\end{equation}
\begin{equation}
\mbox{II. }P\sim \sigma _0+\sigma _1\log s,  \label{eq4}
\end{equation}
\begin{equation}
\mbox{III. }P\sim \sigma _0+\sigma _2\log ^2s,  \label{eq5}
\end{equation}
was performed. Besides the Pomeron non-exchange-degenerate
sub-leading contributions were included. The intercepts of the
$f/a$ and $\omega /\rho$ Regge trajectories were assumed to be
different.

The analysis \cite{Ezhela} is the most complete and professional
one among those existing. It shows e.g. that the lowest energy,
where the above parametrization works (i.e. $\chi ^2/dof\sim 1$)
is $\sqrt{s}=9$ $GeV$.

In the present paper we continue the program initiated by \cite{Ezhela} by
scrutinizing two more earlier unexplored models for the Pomeron. We show
that they do not require the secondary trajectories to be
exchange-degenerate. This disputable point was recently raised also in ref.
\cite{GN}. As seen from Fig. 1 the 10 resonances belonging to the 4
different $I^G(J^{PC})$ families $\rho -\omega -f_2-a_2$ are compatible with
a unique linear exchange-degenerate Regge trajectory with $\alpha (0)=0.48$
intercept.

\begin{center}
\includegraphics*[scale=0.6]{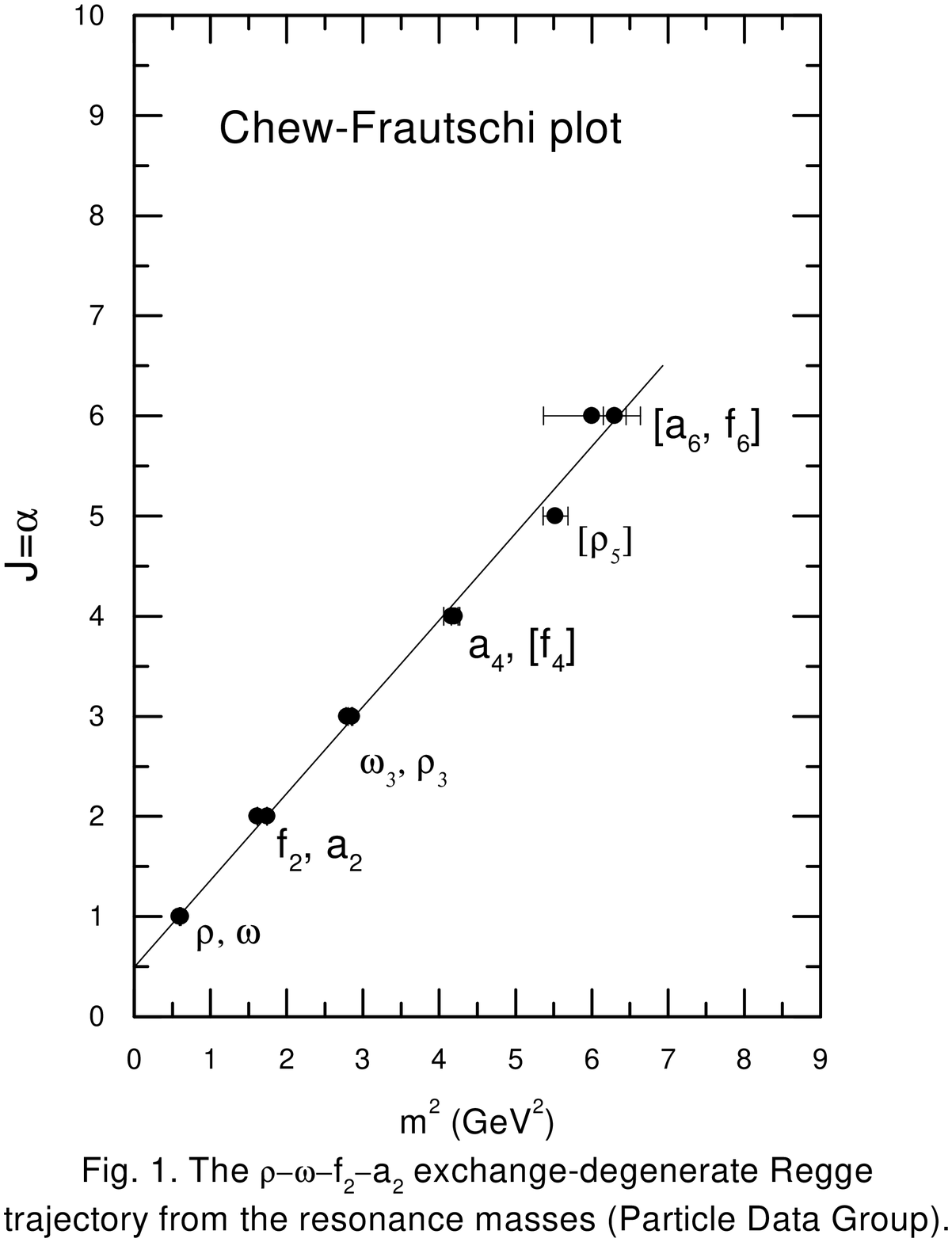}
\end{center}

\smallskip\

Our procedure follows the paper \cite{Ezhela} as close as possible. Firstly,
we use the same data sample \cite{Computer}, secondly, we use the same
number of free parameters (namely, 16), thirdly, we calculate the errors in
the same way as the authors of ref. \cite{Ezhela} do. Moreover, we use the
same notation. This universality makes it possible to confront the resulting
fits without any bias.

\section{Alternating series for the Pomeron}

Let us write the real and imaginary parts of the forward elastic scattering
amplitude as
\begin{equation}
\frac 1sImA_{h_1h_2}\left( s\right) =ImP\left( s\right) +ImR\left( s\right) ,
\label{eq6}
\end{equation}
\begin{equation}
\frac 1sReA_{h_1h_2}\left( s\right) =ReP\left( s\right) +ReR\left( s\right) ,
\label{eq7}
\end{equation}
where
\begin{equation}
ImP\left( s\right) =\lambda _{h_1h_2}\left[ A+B\log \left( \frac
s{s_0}\right) +C\log ^2\left( \frac s{s_0}\right) \right] ,  \label{eq8}
\end{equation}
\begin{equation}
ReP\left( s\right) =\frac \pi 2\lambda _{h_1h_2}B+\pi \lambda _{h_1h_2}C\log
\left( \frac s{s_0}\right) ,  \label{eq9}
\end{equation}
comes from the Pomeron contribution and
\begin{equation}
ImR\left( s\right) =\left[ Y_1^{h_1h_2}\mp Y_2^{h_1h_2}\right] s^{-\eta },
\label{eq10}
\end{equation}
\begin{equation}
ReR\left( s\right) =-\left[ Y_1^{h_1h_2}\cot \left( \frac{1-\eta }2\pi
\right) \pm Y_2^{h_1h_2}\tan \left( \frac{1-\eta }2\pi \right) \right]
s^{-\eta },  \label{eq11}
\end{equation}
corresponds to the contribution of two exchange-degenerate Regge
trajectories as in the classical paper of Donnachie and Landshoff
\cite{DL}. Everybody is aware of the fact that exchange degeneracy
is violated. The question is only the amount of this violation and
the price to be paid for the introduction of additional free
parameters. The quality of our fits show (see below) that exchange
degeneracy is a reasonable approximation to reality.

We have fitted the above model to the data on total cross sections
and the ratio of the real to imaginary part of $pp$,
$\overline{p}p$, $\pi ^{\pm }p$, $K^{\pm }p$, $\gamma p$ and
$\gamma \gamma $ scattering, starting from $\sqrt{s_{\min }}$
until the highest available energies. As in ref. \cite {Ezhela},
we have studied the stability of our fit by varying the lower
limit of the fit $\sqrt{s_{\min }}$ between $3$ and $13$ $GeV$.
The number of the experimental data points and the resulting fit
(value of $\chi ^2/dof$) as well as the dependence of the fitted
parameters on the lower bound $\sqrt{s_{\min }}$ are presented in
Fig. 2.

The results of our fits basically are consistent with those of
ref. \cite {Ezhela}. The minor difference is in the limiting value
$\chi ^2/dof\sim 1$, reached at $\sqrt{s_{\min }}=10$ $GeV$ in our
fits (instead of $9$ of ref. \cite{Ezhela}). The difference
obviously comes from exchange degeneracy of the subleading
trajectories imposed here but relaxed in \cite{Ezhela}. The fits
for the total cross sections and $\rho $-ratio (for $\sqrt{s_{\min
}}=10 $ $GeV$) extrapolated to $\sqrt{s_{\min }}\geq 3$ $GeV$, are
shown in Fig. 3.

An important result of our fit is that the coefficients of the
series (\ref {eq1}) have alternating signs (see Table 1). This
property and the rapid decrease of their absolute values (as $\sim
1/10$) provides the fast convergence of the series and ensures the
applicability of this approximation at still much higher energies.
The physical motivation of such a finite series representation of
the Pomeron was discussed earlier - both in the context of Regge
multipoles \cite{IL,DJ} and QCD \cite{FJKLPP}.

\smallskip\

{\footnotesize
\begin{tabular}{|cccccc|}
\hline \multicolumn{1}{|c|}{$A$ (mb)} & \multicolumn{1}{c|}{$B$
(mb)} & \multicolumn{1}{c|}{$C$ (mb)} & \multicolumn{1}{c|}{$\eta
$} & \multicolumn{1}{c|}{$\chi ^2$/d.o.f.} & statistics \\ \hline
\multicolumn{1}{|c|}{$37.1\pm 1.6$} &
\multicolumn{1}{c|}{$-1.55\pm 0.33$} &
\multicolumn{1}{c|}{$0.275\pm 0.018$} &
\multicolumn{1}{c|}{$0.497\pm 0.019$} &
\multicolumn{1}{c|}{$1.03$} & $314$ \\ \hline
\multicolumn{1}{|c|}{} & \multicolumn{1}{c|}{$pp$} &
\multicolumn{1}{c|}{$ \pi p$} & \multicolumn{1}{c|}{$Kp$} &
\multicolumn{1}{c|}{$\gamma p\times 10^{-3}$} & $\gamma \gamma
\times 10^{-4}$ \\ \hline \multicolumn{1}{|c|}{$\lambda $} &
\multicolumn{1}{c|}{$1$} & \multicolumn{1}{c|}{$0.6248\pm 0.0019$}
& \multicolumn{1}{c|}{$0.5502\pm 0.0029$} &
\multicolumn{1}{c|}{$3.068\pm 0.050$} & $0.0822\pm 0.0059$ \\
\multicolumn{1}{|c|}{$Y_1$ (mb)} & \multicolumn{1}{c|}{$49.5\pm
4.7$} & \multicolumn{1}{c|}{$12.0\pm 2.4$} &
\multicolumn{1}{c|}{$-6.2\pm 2.8$} & \multicolumn{1}{c|}{$44\pm
26$} & $0.5\pm 4.2$ \\ \multicolumn{1}{|c|}{$Y_2$ (mb)} &
\multicolumn{1}{c|}{$25.1\pm 2.9$} & \multicolumn{1}{c|}{$5.72\pm
0.69$} & \multicolumn{1}{c|}{$10.5\pm 1.2$} &
\multicolumn{1}{c|}{} &  \\ \hline
\end{tabular}
}

\smallskip\

Table 1: Values of the fitted parameters in the alternating series
Pomeron model

(6)-(11), with a cut-off $\sqrt{s}=10$ $GeV$ and $s_0=1$ $GeV^2$ fixed.

\section{Two-component Pomeron}

The main virtue of the D-L model of the Pomeron \cite{DL} is its
simplicity. Although it the single-term (factorizable!) D-L model
(appended by a nonleading term) fits the data remarkably well
\cite{DL}, its possible extensions involving more free parameters
are obvious and were discussed by various authors
\cite{Ezhela,CMG}. One possibility is to add a constant term,
resulting in the following expression for the Pomeron contribution
\cite{DGLM,GN}
\begin{equation}
P\sim \sigma _0+\sigma _1s^\varepsilon .  \label{eq12}
\end{equation}
Written as
\begin{equation}
ImP\left( s\right) =\lambda _{h_1h_2}\left[ A+Bs^\delta \right] ,
\label{eq13}
\end{equation}
\begin{equation}
ReP\left( s\right) =-\lambda _{h_1h_2}Bs^\delta \cot \left(
\frac{1+\delta } 2\pi \right) ,  \label{eq14}
\end{equation}
the model has the same number of free parameters as in the previous case
(Section 2) or in ref. \cite{Ezhela}. Similar to the previous case, see eqs.
(\ref{eq10}) and (\ref{eq11}), the model is appended by a pair of
exchange-degenerate sub-leading trajectories.

The optimal value of $\chi ^2/dof\sim 1$, as in the previous case, is
reached when $\sqrt{s_{\min }}=10$ $GeV$.

The resulting fits are shown in Fig. 4. The fits for the total cross
sections and $\rho $- ratio for $\sqrt{s_{\min }}=10$ $GeV$, extrapolated in
the same way as for the previous case (see Section 2), are shown in Fig. 5.

Table 2 quotes the values of the fitted parameters with their
errors calculated with the ''threshold value'' $\sqrt{s_{\min
}}=10$ $GeV$. A pair of exchange-degenerate sub-leading
trajectories was also included.

\smallskip\

{\footnotesize
\begin{tabular}{|cccccc|}
\hline \multicolumn{1}{|c|}{$\epsilon $} & \multicolumn{1}{c|}{$A$
(mb)} & \multicolumn{1}{c|}{$B$ (mb)} & \multicolumn{1}{c|}{$\eta
$} & \multicolumn{1}{c|}{$\chi ^2$/d.o.f.} & statistics \\ \hline
\multicolumn{1}{|c|}{$0.1286\pm 0.0092$} &
\multicolumn{1}{c|}{$18.7\pm 2.3$} & \multicolumn{1}{c|}{$8.3\pm
1.4$} & \multicolumn{1}{c|}{$0.480\pm 0.018$} &
\multicolumn{1}{c|}{$1.09$} & $314$ \\ \hline
\multicolumn{1}{|c|}{} & \multicolumn{1}{c|}{$pp$} &
\multicolumn{1}{c|}{$ \pi p$} & \multicolumn{1}{c|}{$Kp$} &
\multicolumn{1}{c|}{$\gamma p\times 10^{-3}$} & $\gamma \gamma
\times 10^{-4}$ \\ \hline \multicolumn{1}{|c|}{$\lambda $} &
\multicolumn{1}{c|}{$1$} & \multicolumn{1}{c|}{$0.6253\pm 0.0019$}
& \multicolumn{1}{c|}{$0.5514\pm 0.0030$} &
\multicolumn{1}{c|}{$3.068\pm 0.051$} & $0.0822\pm 0.0061$ \\
\multicolumn{1}{|c|}{$Y_1$ (mb)} & \multicolumn{1}{c|}{$64.2\pm
4.0$} & \multicolumn{1}{c|}{$22.8\pm 1.5$} &
\multicolumn{1}{c|}{$4.5\pm 1.8$} & \multicolumn{1}{c|}{$100\pm
22$} & $2.1\pm 3.8$ \\ \multicolumn{1}{|c|}{$Y_2$ (mb)} &
\multicolumn{1}{c|}{$22.5\pm 2.4$} & \multicolumn{1}{c|}{$5.3\pm
0.6$} & \multicolumn{1}{c|}{$9.5\pm 1.0$} & \multicolumn{1}{c|}{}
&  \\ \hline
\end{tabular}
}

\smallskip

Table 2: Values of the fitted parameters in the two-component Pomeron model
(6),

(7), (10), (11), (13) and (14), with a cut-off $\sqrt{s}=10$ $GeV$
and $s_0=1 $ $GeV^2$ fixed.

\section{Conclusions}

We have fitted two hitherto unexplored models for the Pomeron to the data
and compared the results with other fits \cite{Ezhela}. We find that:

1. Exchange degeneracy is a good approximation to reality;

2. The series (\ref{eq1}) has alternating signs;

3. The quality of the models fitted is not inferior to similar ones with the
same number of free parameters.

4. Both $\varepsilon $ and $\eta $ correspond to those explored in \cite{GN}.

The authors are grateful to L. L. Jenkovszky for useful
discussions and support.

\end{document}